%% file: cosmosI.tex
\documentclass[aps, nofootinbib]{revtex4}
\usepackage{graphicx}
\usepackage{hyperref}
\hypersetup{
    pdftitle={The Cosmos in Your Pocket: How Cosmological Science Became Earth Technology.  I.},
    pdfauthor={W.T. Bridgman},     
    pdfpagemode={UseOutlines},
    hyperfootnotes={true}
}

\begin{document}
\setcounter{tocdepth}{1}
\title{The Cosmos in Your Pocket: \\
How Cosmological Science Became Earth Technology.  I.\footnote{This paper grew out of some notes originally outlined in a interview for Robert Lippens' ``Big Bang and Creationism'' podcast series (June 12, 2006) and a presentation in the Science Program track at the Balticon Science Fiction convention (May 31, 2007).}}
\author{W. T. Bridgman}
\affiliation{Global Science \& Technology, Inc.}
\email{william.bridgman@gst.com}

\begin{abstract}
Astronomy provides a laboratory for extreme physics, a window into environments at extremes of distance, temperature and density that often can't be reproduced in Earth laboratories, or at least not right away.  A surprising amount of the science we understand today started out as solutions to problems in astronomy.  Some of this science was key in the development of many technologies which we enjoy today.  This paper describes some of these connections between astronomy and technology and their history.\end{abstract}


\maketitle
\section{Introduction}

Many professional popularizers of astronomy, when asked ``Just what good is it to study astronomy?'', usually respond with very spiritual notions like it helps humankind to explore and understand their place in the universe.  This is a perfectly good reason, but it is far from the whole story.  

The Astronomy FAQ provides a few more reasons but doesn't really go into the details.\footnote{The Astronomy FAQ: What good is astronomy anyway? What has it contributed to society? \url{http://www.faqs.org/faqs/astronomy/faq/part2/section-3.html}}   Astronomy has a little known practical side.  When we work to understand the cosmos and the realm of the universe beyond the Earth, we very quickly find ourselves encountering distance and time scales, as well as ranges of temperature and pressure, far beyond those accessible in Earth-based laboratories.  In a surprising number of cases, we've discovered and explored properties of matter under extreme conditions in astrophysical environments long \textit{before} laboratory technology made it possible to test the properties on Earth.  This fact has been known by professional astronomers for many years, receiving mention in a number of excellent historical overviews such as Plaskett in 1922\citep{1922JRASC..16..137P} to more recently by Longair\citep[pg 3]{2006ccha.book.....L}.

The notion that physical principles on the the Earth also apply to the rest of the cosmos is not a new idea.  Galileo suggested this as far back as 1592.  

I've often heard the complaint, mostly from various pseudo-scientific groups, that we have no way of knowing, or ``proving'', what is really happening `out there' in the distant cosmos.  They also try to make a distinction between what they like to call ``origins'' science vs. ``real'' or ``operational'' science.  This distinction is made so the pseudo-science advocates can claim there is a difference between the science that generates products and the science that makes predictions that are contrary to their particular dogma.  Such statements are then used to justify all kinds of outlandish hypotheses about how the rest of the universe came to be or operates, usually to satisfy some political agenda.  This treats astronomy as some kind of `spectator science', not an avenue of inquiry which can yield important insights into physics we can test on Earth.  Yet the history of science suggests otherwise.

This erroneous notion of our understanding of the universe ignores that fact that while many astrophysical explanations are based on extrapolations of the physics we know on Earth, such extrapolations can also provide feedback that we can investigate in Earth laboratories.  These types of discoveries provide checks on our extrapolations, and provide new knowledge that can have technological applications.

So what is the appropriate response for a teacher to give to the question, ``Why should I care about astronomy?''

The answer is that you should care about astronomy if you wish to live in a society with the benefits of technology.  The development of technology sometimes requires the ability to test the science in new and novel environments.  Sometimes these environments are not available to Earth-based laboratories, but could well be with minor technological advances.  Astrophysics can provide some more extreme environments for these tests.

In this paper, I'll travel along a slightly skewed path from the standard historical treatments of astronomy history.  I'll walk you through many important events along the history of astronomy and astrophysics, with a few detours not only into Earth laboratories, but the Patent Office as well, illuminating some of the interconnections.  We'll explore not only ``what did they know?'', and ``when did they know it?'', but ``What did others do with the knowledge?'' and ``How does it impact my life today?''

Whenever possible, I have attempted to examine the original papers when reporting on key discoveries, rather than relying entirely on historical overviews.  Some of the meta-references I used for overview material and tracking down historical papers are Hernshaw's ``The Analysis of Starlight''\citep{1986asoh.book.....H}, Clayton's ``Principles of Stellar Evolution and Nucleosynthesis''\citep{1983psen.book.....C}, Longair's ``The Cosmic Century: A History of Astrophysics and Cosmology''\citep{2006ccha.book.....L} and van Helden's ``Measuring the Universe: Cosmic Dimensions from Aristarchus to Halley''.  Wikipedia\footnote{\url{http://en.wikipedia.org/wiki/Main\_Page}} provided many helpful pointers to original sources in tracking down details on some of the technologies.  It was particularly invaluable in tracking down connections of changing terminology as science migrated from theory to laboratory phenomena to engineering technologies.

This is the first of a series of papers explaining how the study of astronomy has led to advances in technology.


\input{gravity}

\input{taleofelements}
\input{colorofbinarystars}
\input{atoms2stars}
\input{carbon12}

\input{implications}

\begin{acknowledgments}

First and foremost, I want to thank my M.S. and Ph.D. advisor, Don Clayton, for telling the story of the ${\rm^{12}C}$ resonance in his nuclear astrophysics class.  That story provided the seed for this project.

I'd also like to thank Mary Baxter, Samir Chettri (Global Science \& Technology, Inc.), David Batchelor (NASA/GSFC), and Ernie Wright (UMBC) for reading the drafts and providing feedback for improvement.

This work has made extensive use references and papers through the Smithsonian/NASA Astrophysics Data System (\url{http://www.adsabs.harvard.edu/}).  I'd also like to thank the National Science Foundation (\url{http://www.nsf.gov}) and the National Solar Observatory (\url{http://www.nso.edu}) for their solar spectrum data, the Homer E. Newell Memorial Library at Goddard Space Flight Center for access to their journal collection and the United States Library of Congress.    Participants in the online ``History of Astronomy'' discussion provided clarification on a number of historical details.

I'd also like to thank the many scientists who assisted me with pointers to historical details which saved me weeks of work sifting through citation lists: Craig DeForest (SwRI), Therese Kucera (NASA/GSFC), and Steven Dick (Chief Historian, NASA/HQ) and Ed Salpeter (Cornell University).
\end{acknowledgments}

\raggedright
\bibliographystyle{plainnat}
\bibliography{cosmosI}
\end{document}

%% file: gravity.tex
\section{Gravity: Interpolations to a Small Planet}
\label{sec:gravity}
\subsection{Building a Theory}
Galileo opened the door to the exploration of the Universe through observation and experiment, using his telescope and experiments with falling objects and pendulums.  The next big accomplishment would require a major leap of insight, which would also bring with it a dramatic improvement in the mathematical tools for examining the cosmos.  Today, we think of this accomplishment as obvious, yet according to legend, its originator didn't recognize it until he was hit on the head.

The person in question is Sir Isaac Newton and the theory is his ``Principle of Universal Gravitation'', first published in 1687 in his work, \textit{Principia Mathematica}.  From this work, Newton's most famous equation is:
\begin{equation}
F = {G M m \over {r}^{2}}
\end{equation}
where $F$ represents the force of attraction between two masses, $M$ and $m$, $r$ is the distance between those masses, and $G$ is some constant which Newton didn't know but which was included to make sure $F$ had the appropriate units of force.  Newton claimed that this expression was valid from the Earth into the solar system and possibly beyond, hence the name ``universal''.

To fully understand the scope of this achievement, we must first examine just what was known at the time.  Newton said he saw further by standing on the shoulders of giants.  Let's meet some of the giants.
\begin{itemize}
\item Galileo had established the mass independence of objects falling under the force of gravity around 1590.  There is some evidence that others, such as Giambattista Benedetti, had come to this realization as earlier.\footnote{\href{http://en.wikipedia.org/w/index.php?title=Giambattista_Benedetti&oldid=267052653}{Wikipedia. Giambattista Benedetti Ñ wikipedia, the free encyclopedia, 2009.} [Online; accessed 28-January-2009]},\footnote{\href{http://en.wikipedia.org/w/index.php?title=Galileo_Galilei&oldid=292642534}{Wikipedia. Galileo Galilei Ñ wikipedia, the free encyclopedia, 2009.} [Online; accessed 27-May-2009]}

\item Johannes Kepler had established his three laws of planetary motion between 1609 and 1619.  Gottfried Wendelin subjected Kepler's laws to an additional test on the moons of Jupiter a few years later (1643), so Newton knew that the trajectories of the planets were ellipses.

\item In 1645, Ismael Boulliau demonstrated that an object subjected to an inverse-square central force law would move around the center of force in an ellipse.  This result conveniently meshed with the work of Kepler and Newton had demonstrated this himself\citep[pg 387]{1980nrbi.book.....W}.

\item In 1672, Giovanni Cassini recognized that if he could measure an absolute distance between the Earth and any other planet, then using Kepler's Laws, he could determine the scale of the solar system.  In collaboration with a French astronomer at another location, he managed to measure the parallax of the planet Mars against the background stars and thereby determine the distance from the Earth to the Sun, commonly called the Astronomical Unit,  or AU.  In spite of severe measuring errors, he managed to obtain a value of 87 million miles, less that 10\% short of the modern value of about 93 million miles\citep{1996Sheehan:qy}.  However, a re-examination of Cassini's analysis suggests his technique and measurements were only effective for placing an upper limit on the solar parallax, or a lower limit on the AU\citep[pp. 239-241]{Dick:OceaJoinNavaObse2002}.  Even so, it is an impressive achievement and Cassini's value would be used for many years\citep[pp. 128-143]{1989Helden:uq} .  

\item By 1678, Robert Hooke, a contemporary of Newton, would conclude that gravity had an inverse-square dependence on distance.
\end{itemize}

Leveraging off this knowledge base, in 1684, Newton realized that the inverse-square distance law, combined with a force proportional to the product of the masses would tie these properties together and work consistently with his own force law, $F = m a$.  Newton spent another three years deriving some of the implications of this theory before he published the result in \textit{Principia}.

\begin{figure}[htbp]
\includegraphics[scale=0.6]{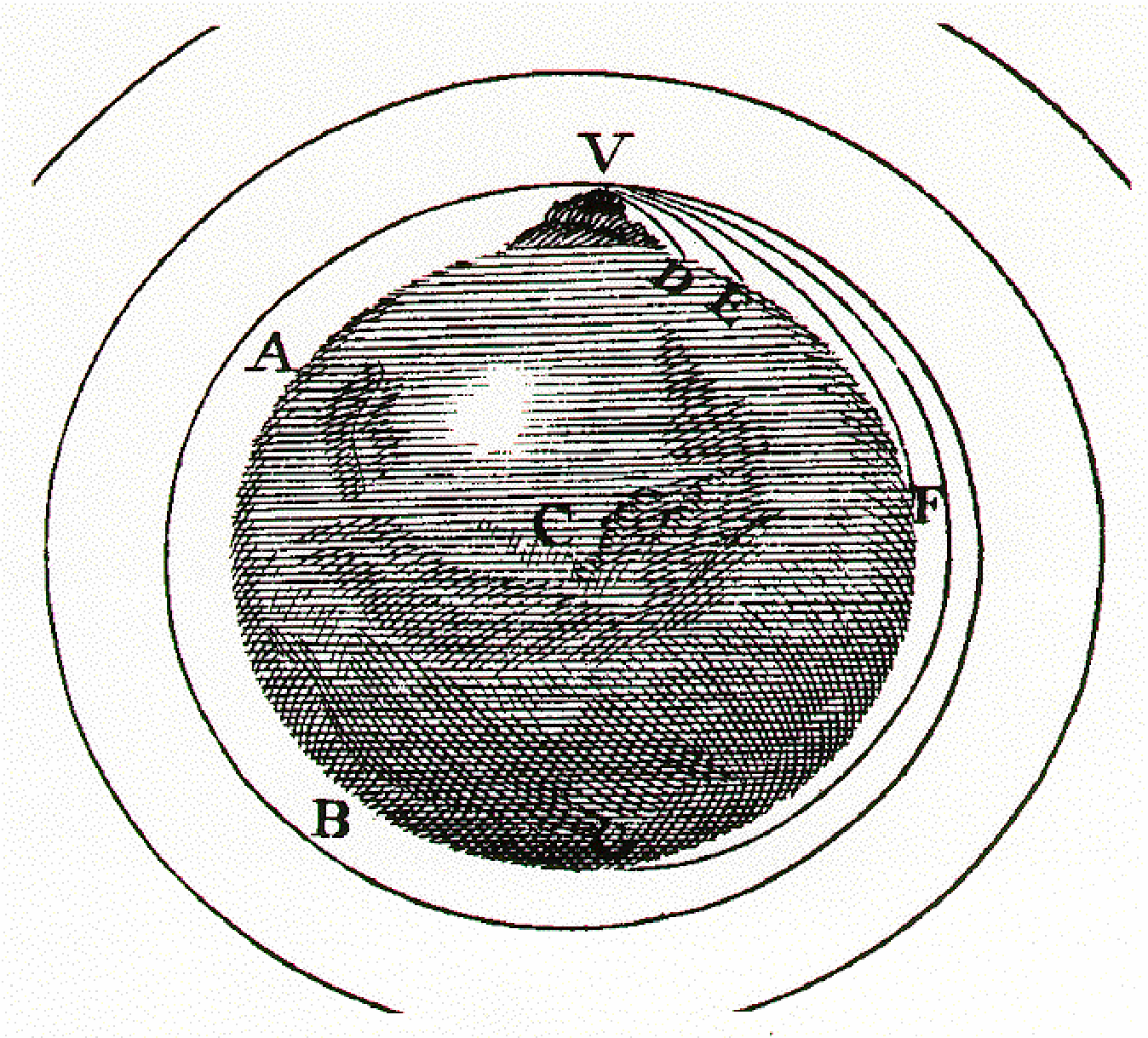}
\caption{Newton's Cannon, the experiment Newton couldn't perform.  As the cannon on the mountain top launches faster projectiles, they manage to travel further around the Earth before striking the ground.  Eventually the projectiles travel fast enough that they miss the curvature of the Earth entirely, achieving orbit.  It would be over 200 years before anyone could actually perform this experiment!  \textit{Image Credit: Michael Fowler, U.Va.}, http://galileoandeinstein.physics.virginia.edu/lectures/newton.html}
\label{fig:newtonscannon}
\end{figure}

Newton understood many implications of this work.  One major prediction was that a projectile, propelled with sufficient velocity tangent to the surface of the Earth, would miss the surface of the Earth and perpetually ``fall'' around it, as illustrated in Figure~\ref{fig:newtonscannon}.  This figure, sometimes referred to as ``Newton's Cannon'' was published in a popularization of the \textit{Principia}, called \textit{A Treatise of the System of the World} (1728).  Unfortunately, no one in Newton's day could perform this experiment!

While this ultimate test of Newton's theory was impossible using the technology of his day, there were plenty of other observational tests which could be performed.

Between 1695 and 1705, Edmund Halley used Newton's work to plot the orbits of comets.  In the process, Halley recognized that several comets seemed to have the same orbit and their years of visibility matched the period from the orbit determination.  He made a prediction for a comet's return but did not live to see the prediction fulfilled.  The comet in question now bears his name, Halley's comet.  

Halley also theorized that a better estimate of the Earth-Sun distance could be obtained by timing the passage of a planet between the Earth and the Sun across the solar disk.  The timing of these passages, called transits, could be converted to precise angular positions against the Sun from which a parallax angle could be derived.  It was expected that Venus, passing closer to the Earth than Mars or Mercury, would yield a larger angle, and would therefore be easier to measure.  Halley also did not live to see this prediction fulfilled.  In 1763, the next transit after Halley's death, international efforts were made to perform the measurements, but they were confounded by the notorious `black drop' effect\footnote{As the disk of the planet approached the limb of the Sun, an `appendage' appears to form, connecting the limb of the Sun to the planet.  This creates an ambiguity in determining the precise time of contact.  This effect is even observed from space-based observatories and the modern interpretation is that it is a diffraction effect due to the wave nature of light\citep{2004Icar..168..249S, 2005MSAIS...6...17L}.} and the results were less accurate than expected.  Nonetheless, scientists managed to estimate a value between 81-98 million miles, consistent with Cassini's result of nearly a century earlier\citep{2004Bell:kx}.


For nearly two hundred years, the experimental verification of Newton's Cannon was beyond the capabilities of engineering technology.  But many researchers would continue to explore the implications of Newton's theory through the power of mathematics and test what they could against observations.  Here are just a few of the highlights of historical and technological significance:

\begin{itemize}
\item \textbf{1772:} Joseph-Louis Lagrange would discover the special solution to what was known as the three-body gravitational problem, which revealed five regions of stability in a system of two massive bodies and one small mass.  The locations would become known as Lagrange Points and designated as L1 through L5.  L1, L2, and L3 would lie along the lines between the two bodies while L4 and L5 would form equilateral triangles with respect to the two large bodies. 

\item \textbf{1774:} Nevil Maskelyne would use the gravitational deflection of a plumb-line by the mass of a mountain in Scotland to estimate the density of the Earth.  This experiment was named the Schiehallion experiment, after the mountain.

\item \textbf{1797:} Henry Cavendish, also seeking to estimate the density of the Earth, conducted an experiment using Newton's principle of gravitation with a torsion balance.  This experiment also provided the first estimate of Newton's Constant, $G$.

\item \textbf{1801:} The early observations of Ceres, the second in a group of minor planets to be found between the orbits of Mars and Jupiter, was interrupted as the Sun passed between the Earth and the object.  Eager to obtain more observations, the mathematician and physicist Carl Fredriech Gauss undertook the task of determining the orbit of the object using the few available observations, to aid in re-acquiring it at a later time.  The effort succeeds the next year, 1802, and makes the 25 year old Gauss a European celebrity\citep{1999MathM...72..83}.

\item \textbf{1803:} William Herschel would be the first to recognize that some stars traveled through space as binary and multiple systems\citep{1997ASPC..130..291S}.

\item \textbf{1821:} Jean Baptiste Joseph Delambre was to the first to recognize deviations in Uranus' orbit from the predicted path.    It would not be until 1841 that John Couch Adams would consider these deviations to be due to an as yet unknown planet and would attempt to use Newton's theory of gravity to predict the location of the unknown planet\citep{1996OConnor:uq}.  The idea that you might detect matter through it's gravitational influence before you could detect it directly, would become a recurring theme in astronomy.  The perturbations of Uranus would become the first ``dark matter'' problem for astronomers.

\item \textbf{1827:} F{\'e}lix Savary does a complete orbit determination of the binary star $\Xi$ Ursa Majoris\citep{1997ASPC..130..291S}.

\item \textbf{1844:} Friedrich W. Bessel discovered anomalies in proper motions (the motion of stars against the background stars) of the nearby stars Sirius and Procyon.  This would become the first extrasolar ``dark matter'' problem.

\item \textbf{1846:} Like the Venus transit eighty years earlier, another case of international competition in science motivates the discovery of the planet Neptune, based on the predications of Frenchman Urbain LeVerrier and Englishman John Couch Adams.  Their calculations are based on the perturbations of the planet Uranus.

\item \textbf{1849:} Antoine Yvon Villarceau would develop an algorithm for determining the real Keplerian orbital elements for binary stars\citep{1997ASPC..130..291S}.  This capability would be the earliest rigorous mathematical indicator that Newton's gravitational force law applied beyond the solar system.

\item \textbf{1859:} Urbain LeVerrier continued his studies of planetary motions and eventually reported that the planet Mercury had an additional shift in it's orbit which could not be explained by perturbations from the other planets.  Success with using such analyses to discover Neptune prompted the search for another planet between Mercury and the Sun, with the suggested name of `Vulcan'\label{LeVerrierMercury}.  In spite of extensive searches, no planet would ever be discovered based on these observations and the solution to this mystery would await the development of Einstein's General Theory of Relativity\citep{2009Bridgman:mz}.

\item \textbf{1862:} Alvan Clark reported the detection of a faint star near Sirius, believed to be a companion of the star\citep{1862Clark:vn}.   This would explain Bessel's 1844 observations.

\item \textbf{1896:} J. Schaeberle discovered a faint companion of the star Procyon, estimating its mass to be about 1/5 that of Procyon and suspecting that was the source of the perturbations reported by Bessel\citep{1896PASP....8..314S, 1896AJ.....17...37S}.  These small, faint stars would prove to harbor a few mysteries of their own that would provide a new laboratory for testing extremes in atomic physics\citep{2009Bridgman:mz}.  The success of these early ``dark matter'' searches would provide an incentive for modern cosmological dark matter searches.

\item \textbf{1906:} An asteroid is discovered in Jupiter's orbit near one of the points of stability first theorized by Lagrange.  It would be the first asteroid discovered at a Lagrange Point.

\item \textbf{1957:} Over 200 years after it was originally proposed, as part of a more hostile international competition, the ``Newton's Cannon'' experiment would finally be performed with the launch of the Sputnik satellite into Earth orbit by the Soviet Union.
\end{itemize}

It is interesting to think about the nature of these achievements based on our understanding of gravity.  For more than 250 years after Newton proposed his theory of gravity, no human could conduct \textit{in situ} experiments to test it.  All the direct measurements possible in Newton's day took place in a thin layer of atmosphere about a mile thick on a sphere about 12,000 kilometers in diameter, and they were extrapolating out to scales over a million times larger.  Ballistic rockets were tested for over a decade prior to Sputnik and there was significant interest by the military for precision impact determination.  However, uncertainties in atmospheric drag made it difficult to determine the contribution from variation of the Earth's gravitational field and identify the Newtonian force law\footnote{David DeVorkin, 2011.  Private communication. American Institute of Physics, July 29, 2011.}.  Even with these limitations, the numbers they obtained for the gravitational field were within 10\% of modern values.  The Astronomical Unit would be refined by radar in 1959, and is today used to navigate interplanetary spacecraft with extraordinary precision using the same principles that not only established the scale of the solar system, but also became the first rung in the cosmic distance ladder for scaling the Universe.

\subsection{Gravity in the Laboratory}

One could make the case that our modern theory of gravity was validated as a laboratory science in 1797 when Henry Cavendish obtained the first experimental measurement of the gravitational constant, G.  However, this feat was accomplished under the \textit{assumption} of the validity of the inverse-square law with distance.  Numerous subsequent efforts to measure G worked under this same assumption.  Nearly a hundred years later, \citet{1895PhRvI...2..321M} reported to have tested the inverse-square law to a precision of 0.2\% at scales between 3.6 to 7.3 centimeters\citep[pg 56]{1987thyg.book...49C}.  It was not until the 1970s that researchers examined these earlier determinations of G to examine the assumption of the inverse-square law in more detail.  To the surprise of many, they found that the earlier measurements suggested that the inverse-square law did \textit{not} apply at laboratory distance scales under about one meter!\citep{1974PhRvD...9..850L}.

There are a number of complexities involved in studying gravity at laboratory scales on the Earth, the first being the weakness of the gravitational force itself makes it difficult to measure.  The second problem is that you can't shield your experiment from the gravitational forces of objects outside the domain of the measurement, including effects of the masses of the measuring equipment itself.  The surprising results of \citet{1974PhRvD...9..850L} opened the door to possible new physics at laboratory scales and inspired many researchers to develop techniques to improve our ability to measure the gravitational attraction\citep{1979PhRvD..19.2320P, 0026-1394-24-S-001}.

But for nearly two more decades, attempts to refine the measurements produced ambiguous results.  Results from different experiments varied over ranges up to forty times larger than the error estimates of the individual measurements, a characteristic that generally suggests some unknown systematic error in the measurement technique\citep{JoshuaP.Schwarz12181998}.  But these uncertainties also opened the door to speculation of a new particle interaction, a ``Fifth Force'', in addition to the four interactions already known to science (gravity, weak nuclear, strong nuclear, and electromagnetic)\citep{fischbach-1996}.

Researchers improved their measurement techniques, trying to understand the discrepancy, since the idea of a new fundamental force had exciting possibilities for research.  However, as the techniques were improved, the measurements eventually converged to the result that a Newtonian inverse-square force law applied at laboratory scales.  As of this writing, Newtonian gravity has been confirmed to scales less than one millimeter\citep{Reynaud:2005xz}.  However, research continues to push this limit to still smaller scales, as deviations from inverse-square behavior would signal the possibility of an additional short-range force, perhaps driven by additional dimensions.  There are even some theories that propose the field responsible for the Cosmological Constant may be detectable on these smaller scales as well\citep{Long:2003ta}.  

\subsection{Newton's Gravity in Today's Technology}

Today, every technology that depends on a satellite (a device which itself integrates many other scientific principles) uses the knowledge of gravitational principles first established by Isaac Newton.  Examples include:  Weather forecasting; Portable phone technology; Audio and Video Communications by satellite; and Space Exploration.  Can you imagine how different your life today would be without any one of these?  

Today, the problem of determining an orbit based on two or three observations, known as the Gauss Problem, is regularly used by the North American Aerospace Defense Command (NORAD) to track objects in orbit\citep[Chapter 5]{1971Bate}.

As we send interplanetary spacecraft to explore the other objects in the Solar System, we find their paths of travel are consistent with distances calculated pre-spaceflight.  Elaborate trajectories are computed to guide spacecraft to distant parts of the solar system using maneuvers called ``gravity slingshots'' around planets.  Spacecraft are also sent to ``park'' near Lagrange points, regions in space where gravitational forces from the Sun and Earth are almost perfectly balanced.  

The Lagrange points, predicted in 1772, today host a number of spacecraft.  The SOlar Heliospheric Observatory (SOHO)\footnote{\url{http://sohowww.nascom.nasa.gov/}} has been positioned at L1, located on the line between the Earth and the Sun, in what is referred to as a `halo orbit', for over a decade now, along with several other lesser-known satellites.  It provides a continous view of the Sun vital for space weather forecasting.  The Wilkinson Microwave Anisotropy Probe (WMAP)\footnote{\url{http://map.gsfc.nasa.gov/}} is positioned at L2, on the Earth-Sun line behind the Earth.  In the future, it will be joined by the James Webb Space Telescope\footnote{\url{http://www.jwst.nasa.gov/}}.

While the engineering achievements are modern, the mathematics and physics that successfully navigated us to the Moon, and more recently the planets, is almost 300 years old.

\subsection{An Alternative History of Gravity?}

A number of pseudo-scientific views hold that if you can't do certain key experiments, then the theory cannot be considered as `proved'.  Some of them even go so far as to hold the theory of gravity as an example of a well established theory.  

Yet as we've seen here, Newton's theory of gravity spent over two hundred years with a status of `unproved' by this criteria.  Today, we use technologies derived from these theories without ever having to think about the convergence of sciences that made it possible.  One thing we can be certain of is that societies that used such a criteria for establishing their science were not among the first to launch satellites into orbit, or to receive the benefits from those satellites.

%% file: taleofelements.tex
\section{Atomic Insights from Cosmic Observations}
\label{sec:spectra}
\subsection{The Dawn of Spectroscopy}
In 1802, William Wollaston was experimenting with prisms and sunlight and noticed that dark lines cut across the bands of color produced when sunlight passed through the prism.  

It would be over a decade later, in 1814, when Joseph von Fraunh{\"o}fer would observe these dark lines and subject them to a more systematic study, recording their positions and intensities.  Fraunh{\"o}fer would designate the dark lines with upper case letters and the fainter lines with lower case.  Many of Fraunhofer's designations survive today in the nomenclature of astrophysical spectra, combining the chemical element with the Fraunh{\"o}fer letter (calcium K, sodium D, etc.)

\begin{figure}[htbp]
\includegraphics[scale=0.6]{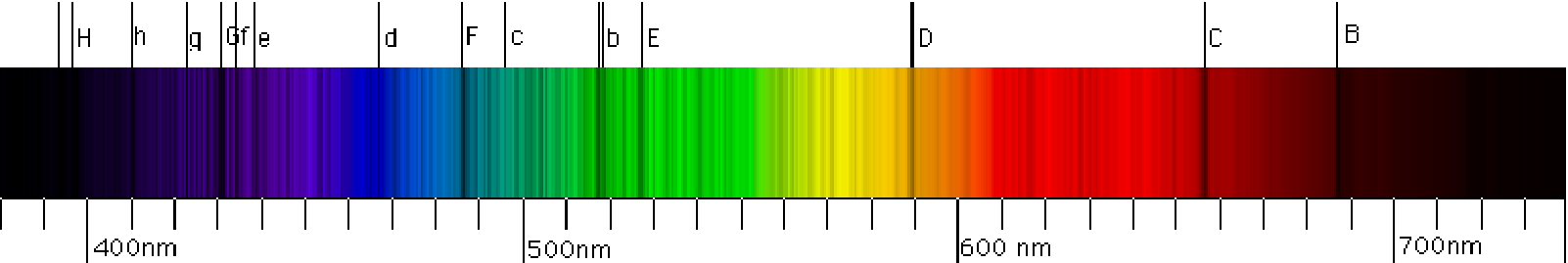}
\caption{A sample solar spectrum generated from modern spectrograph data.  The rainbow distribution of colors is characteristic of continuum spectra.  The dark lines in the color bands are now known as absorption lines.  The lower edge of the image designates with wavelength in nanometers (nm).  Along the top of the image are the upper and lower case alphabetical designations assigned by Fraunh{\"o}fer.}
\label{fig:fraunhofer}
\end{figure}

For years it had been believed that we could never know the composition of distant stars.  In 1835, Auguste Comte suggested that nature of stars could never be known, but even as he was claiming this, the emerging science of spectroscopy was promising to change that.  As early as 1823, William Herschel was suggesting that one could identify the chemical elements by the emitted spectrum\citep[pg 30]{1986asoh.book.....H}.     

The Sun, the brightest of the objects available for study by this new technique, yielded many of its secrets quickly.   By 1860, Robert W. Bunsen and Gustav Robert Kirchhoff, established that all known chemical elements seemed to have a unique signature of lines in its spectrum and identified many of these in the Sun.  Very quickly, astronomers tried these new tools on every object they could see in the telescope.

W.H. Huggins was a pioneer in using spectroscopy to analyze the chemical composition of the stars\citep{1864RSPT..154..413H}, planets and fainter objects, including the nebulae.  While the Sun displayed dark lines (called an absorption spectrum) against a bright rainbow background (called the continuum),  some nebulae displayed bright color  lines against a dark background (called an emission spectrum)\footnote{Note that at this time, the 1860s, the nature of the nebulae was still a mystery, and the physical distinction between galaxies and planetary nebulae as yet undiscovered.}.  

\subsection{Distant Mysteries through the Spectroscope}
In 1864, after pointing his spectroscope at a number of stars, Huggins initiated a study of the nebulae.  This survey included such telescopic favorites as the Cat's Eye nebula in Draco (NGC 6543) and the Dumbbell nebula in Vulpecula, (M27).  He reported many nebulae spectra, specifically planetary nebulae, were radically different from stellar spectra.  Instead of a rainbow continuum spectrum with absorption lines, as seen for the Sun and other stars, he observed an emission-line spectra with bright lines at 500.7nm\footnote{In this work, I'll use the more modern measurement unit of nanometers ($10^{-9}$ meters) instead of the older \AA ngstrom ($10^{-10}$ meters) convention.} and 495.9nm as well as the hydrogen spectral lines designated $\beta$ and $\gamma$.   Huggins initially believed the bright green line seen in many planetary nebulae indicated the presence of nitrogen.  Huggins noted that the spectra of the planetary nebulae were so different from other types of nebulae that they could not be composed of stars\citep{1864RSPT..154..437H} and concluded they were composed of luminous gas.  A later, more refined measurement of this wavelength demonstrated the green `nitrogen' line at 500.7nm  could not be identified with any element known at the time.  

\begin{figure}[htbp]
\includegraphics[scale=0.6]{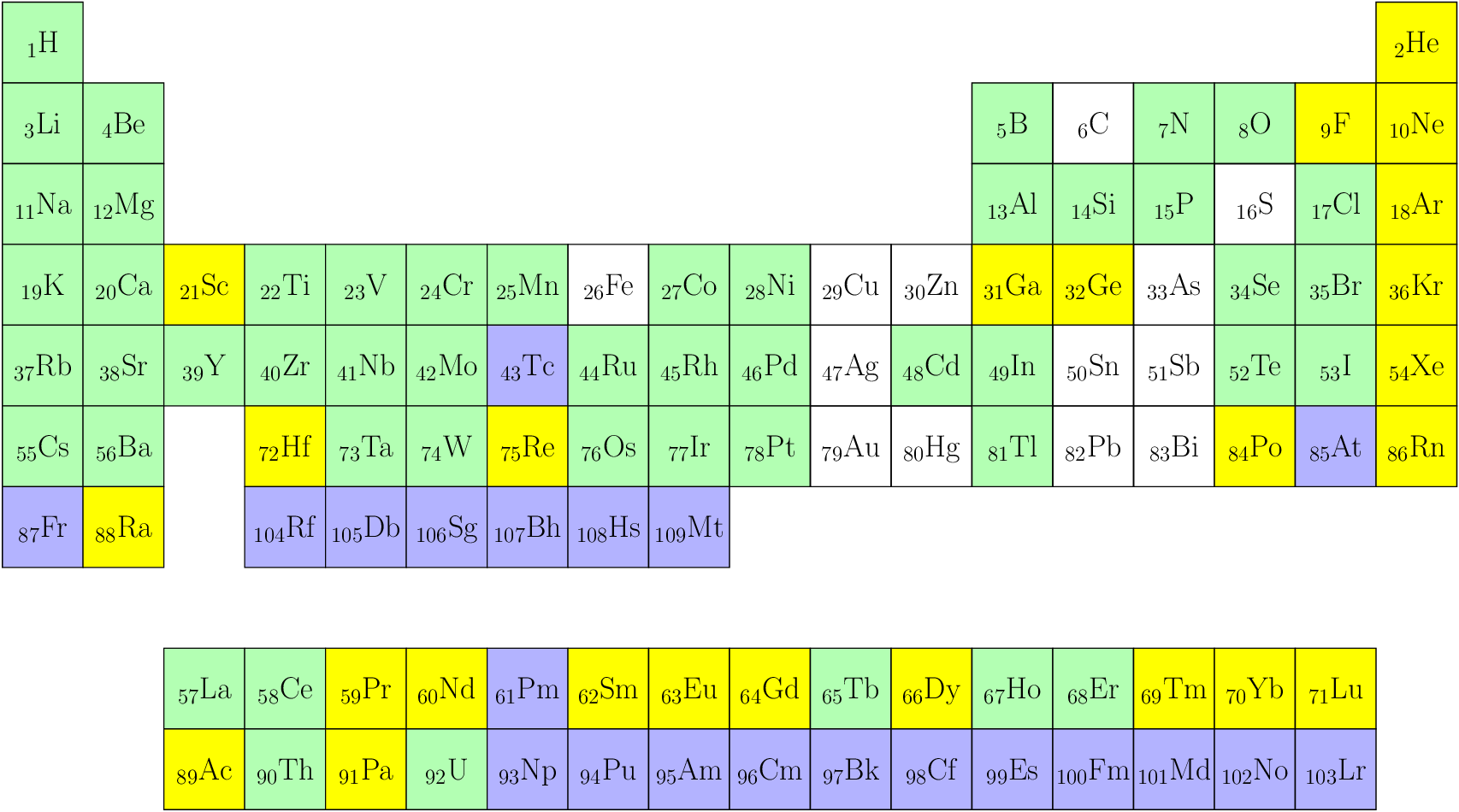}
\caption{A view of changes in the modern periodic table of the elements.  Elements in the white boxes were known in ancient times.  Elements in light green boxes were modern elements identified prior to 1870 (and constituents of Mendelev's first periodic table).  Note that none of the Noble Gases: helium, argon, krypton, etc. were known at this time.  Elements in yellow boxes were discovered between 1870 and 1927, known at the time of Ira Bowen.  From this pattern, it was clear that nebulium could not be an unknown element of low atomic number.  Elements in the light blue boxes were discovered after 1927.}
\label{fig:elementhistory}
\end{figure}

In August of 1868, Pierre Janssen\citep{1869AReg....7..107J} and J. Norman Lockyer\citep{1869Lockyer:qy}  were working independently, but observing the same solar eclipse and experimenting with techniques to observe the limb of the Sun.  In observing the solar prominences, they also observed spectral lines that did not match any known element.  Somewhat boldly, they suggested the lines were due to an as yet undiscovered element which they called helium (from Helios, Greek for `Sun').  Two years later, Dmitri Ivanovich Mendeleev published his first periodic table of the elements in which elements were grouped by their chemical properties.  The structure of Mendeleev's table revealed patterns and gaps, suggesting that some elements were as yet unknown (See Figure~\ref{fig:elementhistory})\citep{2008Pasachoff:sf}.
 
The following year, Charles Young and William Harkness were observing the 1869 solar eclipse for the U.S. Naval observatory\citep[pg 199-200]{Dick:OceaJoinNavaObse2002}.  One of the primary goals of the observations was to search for planets within the orbit of Mercury, these planets which were suggested to exist by the work of LeVerrier(see Section~\ref{LeVerrierMercury}).   Instead, they find an unidentified bright green emission line in the solar corona at 530.3nm.  This line would later be attributed to the hypothetical element, `coronium'\footnote{Many of the papers from this time describe the spectral line as a ``green emission'' or ``green emanation''.  I have wondered if all these discoveries are responsible for the number of green alien substances in comics and science fiction of this era.  Kryptonite is the most notable one that comes to mind, but I suspect there are more.}\footnote{Coronium would make an appearance as a fictional substance in a pair of 1930's science fiction novels, ``The Black Star Passes''\citep{1930Campbell:uq} and ``Islands of Space''\citep{1930Campbell:fk} by John W. Campbell.}.

Twenty-seven years later, in 1895, Sir William Ramsey successfully isolated helium as a gas from uranium ore.  The helium was the product of radioactive alpha decays which would bind with any ambient electrons to form a neutral atom\citep{1895Ramsay:uq}.  He would send a sample of the gas to Lockyer for confirmation.  Unfortunately, the original sample would prove to be unusable, but Lockyer was able to extract more gas for testing and confirm the result\citep{1895Lockyer:fk}.

But there were still new mysteries to be found in the spectra of distant stars.   In 1885, Johann Balmer of Switzerland had discovered a pattern in the laboratory spectra of hydrogen.  The relationship
\begin{equation}
\lambda = 3645.6\times10^{-7} \left({m^2 \over {m^2-n^2}}\right) millimeters
\label{eq:balmer1}
\end{equation}
reproduced the wavelengths of the four prominent visible spectral lines of hydrogen when $n=2$ and $m=3, 4, 5, 6$.  It became known as the Balmer formula.  Was the agreement a coincidence, or a hint at the inner properties of atoms?

In 1896, Edward C. Pickering would report six absorption lines in the spectra of the star $\zeta$ Puppis that did not match any known element.  Initially, Pickering speculated that the lines were due to an element unknown on the Earth\citep{1896HarCi..12....1P}, but he also noticed that his unidentified spectral lines were spaced in a pattern which could be reproduced by a modified form of the Balmer formula, found by replacing $m$ with $m+{1 \over 2}$, or 
\begin{equation}
\lambda = 3645.6\times10^{-7} \left[{(m+{1 \over 2})^2 \over {(m+{1 \over 2})^2-n^2}}\right] millimeters. 
\label{eq:pickering1}
\end{equation}
This similarity with the Balmer formula would subsequently persuade Pickering to attribute these lines to hydrogen\citep{1897HarCi..16....1P}.

In 1898, following the example set by Janssen and Lockyer,  Margaret Huggins (wife of William Huggins), suggested the unidentified nebular line was also due to a new element and proposed, among others, to name it nebulium\citep{1898ApJ.....8R..54H}. 

But progress in isolating nebulium was slow.  There was still no workable theory for explaining the spectra and structure of atoms.  With J.J. Thomson's discovery of the electron as a constituent of atoms in 1897, he developed what became known as the ``plumb-pudding'' model of the atom, where the electrons were embedded in a positively-charged `pudding'.  But in 1911, Ernest Rutherford's experiments indicated that the atom consisted of a dense, positively-charged nucleus much smaller than the atom itself, ruling out Thomson's model.  

From 1911 to 1918, John William Nicholson made several attempts to determine the properties of nebulium\citep{1911MNRAS..72...49N} and even coronium\citep{1913Obs....36..103N} using his model of atomic structure, a variant of J.J. Thomson's ``plumb-pudding'' atomic model\citep{1975McCormmach:vn}.  However, his model generated unreliable and sometimes bizarre results.  His 1918 paper\citep{1918MNRAS..78..349N} goes through fourteen pages of mathematics to obtain an atomic weight for nebulium of 1.31.  To be fair, no one knew details about the structure of the atom at this time, so Nicholson's result did not seem as strange as it does to us today, with our modern knowledge of the atom.  

In the meantime, work also continued on the laboratory identification of the hydrogen lines observed by Pickering.  The lines were finally detected in the laboratory in 1912 by \citet{1912MNRAS..73...62F} in a discharge tube containing hydrogen, but also contaminated with helium.  The lines would still be identified as due to the hydrogen in the tube.  

Then came Niels Bohr's insight in applying quantum principles to the spectrum of the hydrogen atom\citep{1913Bohr:uq}.  One of the key consequences of the Bohr model was the realization that the terms and constants in the Balmer formula (equation~\ref{eq:balmer1}), and more generally the Rydberg formula, for hydrogen spectra came from more fundamental quantities.   By reworking the Rydberg formula in terms of inverse wavelength, ${1 / \lambda} = {E / {h c}}$, Bohr found
\begin{equation}
{1 \over \lambda} = {\pi^2 m_e e^4 \over {8 c h^3 {\epsilon_0}^2}} \left({{1 \over m^2 } - {1 \over n^2}}\right)
\label{eq:hydrogenic}
\end{equation}
where $m_e$ is the electron mass, $e$ is the electron charge, $c$ is the speed of light, $h$ is Planck's constant, and $\epsilon_0$ is the permeability of free space.  It was quickly recognized that the Bohr model could explain the Pickering series, not as a signature of hydrogen, but of singly-ionized helium\footnote{Mathematically, singly-ionized helium behaved like a hydrogen atom with twice the charge and four times the mass.  This general behavior is true of all single-electron or \textit{hydrogenic} ions.}.  By generalizing the equation to nuclei with more protons than hydrogen, $Z>1$, and replacing the electron mass, $m_e$ by the \textit{effective electron mass}, $\mu_e$, which included the fact that the nucleus was not infinitely heavy, the equation for the Pickering series (equation~\ref{eq:pickering1}), was found to be a special case of a more general form of the Rydberg formula
\begin{equation}
{1 \over \lambda} = {\pi^2 \mu_e Z^2 e^4 \over {8 c h^3 {\epsilon_0}^2}} \left({{1 \over m^2 } - {1 \over n^2}}\right)
\end{equation}
This hypothesis received laboratory confirmation, without the hydrogen contaminant, a few years later by \citet{1916AnP...355..901P}\footnote{An historical overview of these spectral components, which was known as the Pickering series, is available in \citet{1922JRASC..16..137P}.  The introduction of Plaskett's paper outlines the importance of astronomical observations in exploring fundamental physics.}  This result helped establish Bohr's quantized orbital model as one of the major successes from this time period.

\subsection{Not So Alien After All...}
In the 1920s, there were few remaining gaps in the periodic table, and none at low atomic weights, where many elements already identified in nebulae resided (see Figure~\ref{fig:elementhistory}).  In 1924, \citet{1924Rosseland:nr} would propose nebulium was due to a metastable atomic state in helium.  A couple of years later, Henry Norris Russell would suggest in his book \textit{Astronomy} that `nebulium' 
\begin{quote}
``must be due not to atoms of unknown kinds but to atoms of known kinds shining under unfamiliar conditions.''\citep[p. 837]{1926QB43.R85a}\end{quote}
These densities would be lower than possible to achieve in the laboratories of the day.  

Ira Bowen, an experimentalist working with Robert Millikan, read Russell's description and realized that meta-stable atomic states already detected in the laboratory might, in conditions of extremely low density, generate transitions which could produce the nebulium spectral lines.  Because these transitions between meta-stable states did not occur under usual laboratory conditions, they were labeled `forbidden'.  There were also theoretical reasons to label these transitions as forbidden, since such a transition would violate a Selection Rule from quantum theory, which kept angular momentum conserved\footnote{Selection rules are properties between initial and final atomic states that determine if a transition can take place between those states.  They are usually constrained by conservation laws, such as angular momentum.}.  These transitions weren't really forbidden in an absolute sense, for there were other mechanisms by which the transitions could occur, that were usually referred to as higher-order electromagnetic multipole transitions.  But making the transition by these other mechanisms had much lower probability per unit time of occurring.  This lower probability per unit time of transition is equivalent to increasing the mean lifetime of the state.  Bowen computed spectral line wavelengths for transitions between several meta-stable states known in oxygen and nitrogen\citep{1979Hirsh:vn}.  Many of the resulting wavelengths matched the nebulium lines and he published these results in a series of papers between 1927 and 1928\citep{1927PASP...39..295B,1927Bowen:uq,1928ApJ....67....1B}.

A few years later, in 1931, two of the `nebulium' lines (oxygen at 630.0nm  and 636.4nm) were reproduced in the laboratory\citep{PhysRev.37.160}.  It took a two hour photographic exposure to record the faint emission line from a quartz discharge tube operating at a pressure less than 0.003 atmospheres.  Over the decades of the 1930s through 1940s, the meta-stable transitions were established in a firm theoretical framework through quantum mechanics.  This success became yet another validation of the idea that, even at the atomic level, physics was the same throughout the cosmos.

The solution to the coronium question would not be found until 1942.  Walter Grotrian pointed out that energy levels in iron that had lost nine (Fe$^{+9}$ or Fe X) and ten (Fe$^{+10}$ or Fe XI) electrons had energy differences corresponding to the coronium lines at 637.4nm and 789.2nm\citep{1943ApJ....98..116S}\footnote{For various historical reasons, some astronomical papers use a rather archaic notation, combining the chemical symbol with a roman numeral.  In this system, neutral atoms are designated with the Roman number I, so neutral hydrogen, helium and carbon, would be designated HI, HeI, and CI, respectively.  Ions are designated with one higher Roman numeral, so ionized hydrogen is HII and ionized helium is HeII, and so on.  Chemists and physicists prefer the superscript notation where HII would be written H$^{+1}$ and HeII is HII = He$^{+1}$ }.  Motivated by this information, Edl\'en used isoelectronic sequences\footnote{Isoelectronic sequences are atoms of different atomic numbers (protons) with the same number of electrons, usually comparing atoms with ions.  For example, singly ionized helium, He$^{+1}$, (Z=2) is isoelectronic to the hydrogen atom (atomic number, Z =1), and to doubly-ionized lithium, Li$^{+2}$, (Z=3).  These sequences are especially useful in the understanding the energy level structure of multi-electron atoms since they differ only by the nuclear charge in the center.} to explain coronium as forbidden transitions in high ionization states of iron (Fe$^{+12}$ or Fe XIII), nickel and calcium\citep{2006:coronium,1942Elden:fj}.  

Again, we find that astronomy provided a laboratory of extreme physics in cases where laboratory science was not yet up to the challenge.  In atomic physics, it provided us with not only the hint of a new element (helium), but a test of unusual states in atomic physics at high temperatures and low densities which would take a few more years to reproduce in Earth laboratories.  The misidentification of these states is not that unusual.  Many other elements were hypothesized in the early days of atomic physics to explain anomalous observations\citep{1997Perez-Bustamante:lr}.  Today, only helium survives as an actual new entry in the periodic table.

\subsection{The Cosmic Impact on the Understanding of Atomic Structure}
Just over one hundred years ago, helium, once discovered, was still an exotic element, difficult to extract from the Earth.  Its utility caused that to change quickly.  By 1911, liquid helium was being used as coolant for the first mercury superconductor\citep{1994AmJPh..62.1105M, deBruynOuboter:1997:HKO}.  Its low atomic weight and quantum properties also made it the first discovered superfluid  in 1937\citep{1938Kapitza:sf, 1995PhT....48g..30D}.  Today, superfluids are used in precision devices as a working medium when precision parts need to work together frictionlessly.

Today, atomic ``forbidden'' lines are utilized to measure temperature and density in low-density plasmas, specifically in controlled fusion experiments\citep{1973ApJ...183L..43F,1998Trabert:kx,2003HyInt.146..269T,2005AIPC..774...33W}.  These types of atomic transitions are studied in detail to analyze x-ray observations of black holes and other astrophysical sources seen by space-based observatories such as Chandra and ASTRO-E2\citep{2005AIPC..774...83B,2005AIPC..774..155D}.  Modern spectral simulation codes such as XSTAR\citep{2007Kallman:fk, 2001ApJS..134..139B}, CLOUDY\citep{1998PASP..110..761F}, SPEX\citep{1996uxsa.coll..411K}, and CHIANTI\citep{1998ASPC..143..390D} are used to model both astrophysical and laboratory plasmas and provide feedback on improving atomic structure models.  Work making laboratory identifications of spectral lines in astrophysical sources, especially high energy sources like AGN/quasars, black holes, and neutron stars, is ongoing.  This work is used not only to test our understanding of these distant objects, but also to test our understanding of atomic properties at extreme temperatures and densities\citep{2005AIPC..774..155D}.  While I've yet to find a clear example of a technology dependent on some of the more easily produced ``forbidden lines'', they do have an impact on the development of atomic modeling techniques.  Many of these techniques are incorporated into software for ``designing'' molecules with unique properties for pharmaceuticals or materials science which finds its way indirectly into other technologies and products.\footnote{\href{http://en.wikipedia.org/w/index.php?title=Computational_chemistry&oldid=292886722}{Wikipedia. Computational chemistry Ñ wikipedia, the free encyclopedia, 2009}. [Online; accessed 28-May-2009]}

\subsection{A final note: Geocoronium}
Back in 1869, during the initial wave of discoveries from spectroscopes pointed skyward, Anders \AA ngstrom pointed a spectroscope at the aurora and detected several lines, the brightest of which was a green line near 557.7 nm\citep{1937RvMP....9..403H}.  Some years later, about 1912, with the source of the line still unidentified, Alfred Wegener (of continental drift fame) would propose the name `geocoronium' as a new element for the source of the line\citep{1912Wegener:lr}.  This name never seemed to obtain any kind of wide use, probably because by then the periodic table was sufficiently well understood that the probability of another undiscovered element, hiding among the known gases in the Earth's atmosphere, was regarded as very low.  Most efforts concentrated on identifying the line in the spectra of known gases.  The lines would later be identified as atomic oxygen in the Earth's upper atmosphere by McLennan and collaborators\citep{1925PRSLA..108..501M, 1927PRSLA..114....1M}.  

%% file: colorofbinarystars.tex
\section{The Color of Binary Stars}
\subsection{A Logical Explanation}
Into the 1800s, the growth of interest in astronomy by academics and amateurs continued.   There was an explosive growth in data collection, by pen and paper, as more and more telescopes were pointed into the night sky.  In 1802, William Herschel, who recorded the positions of many stars, would conclude that a number of stellar pairings observed in the heavens (called optical doubles) were indeed gravitationally bound (sometimes referred to as visual doubles).  These binary stars became a new area of interest for astronomers, in part because it was recognized that Kepler's Laws might provide a means of determining the actual masses of distant stars.  

One particularly interesting pattern observed among double stars were the large number of pairings with dramatically different colors.  One of the favorite pairings familiar to many amateur astronomers is $\beta$ Cygni, also known as Albireo, which consisted of a bright orange star and a fainter blue companion.

A possible explanation for the ubiquity of these types of pairings was proposed at the Royal Bohemian Scientific Society, on May 25, 1842.  Under the title  \textit{``Concerning the coloured light of double stars and of some other heavenly bodies''},
Johann Christian Doppler (1803-1853) proposed that the underlying cause of the colors of these pairs was their orbital velocities relative to the observer on Earth.  

\begin{figure}[htbp]
\includegraphics[scale=0.8]{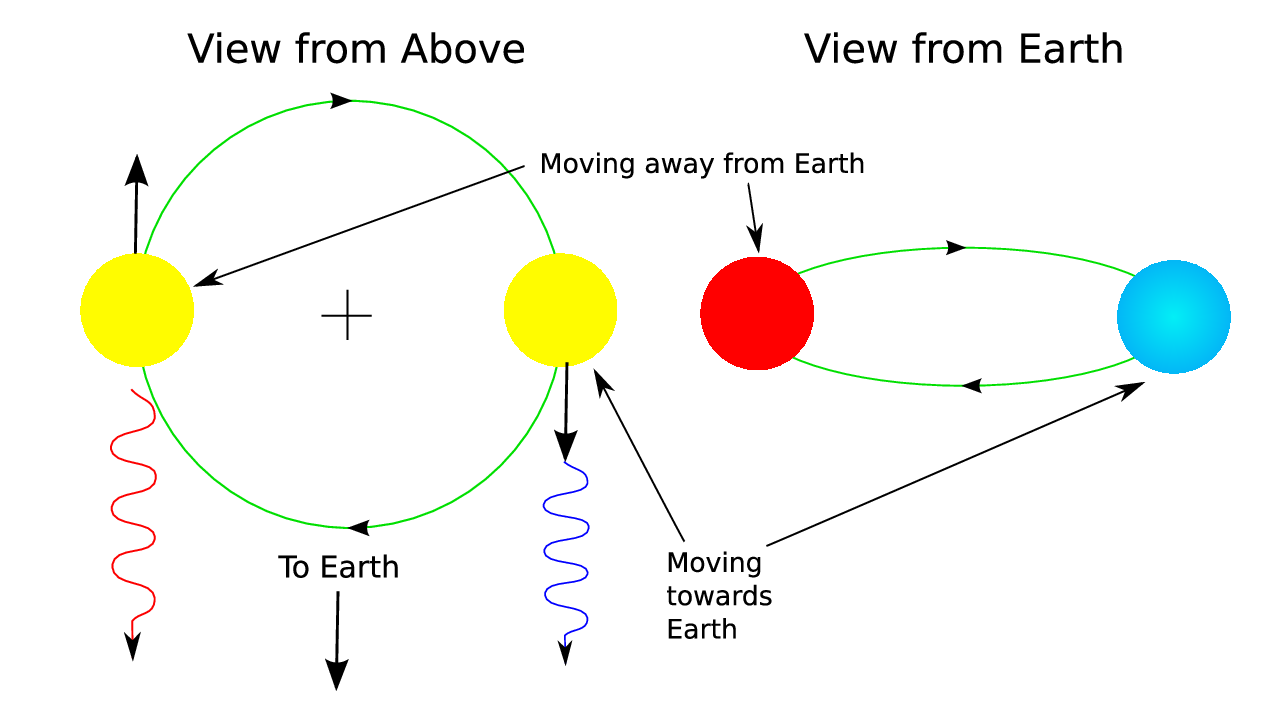}
\caption{Two views of a binary star system illustrating Doppler's original proposal.}
\label{fig:doppler}
\end{figure}

In the same year as the discovery of binary stars, Thomas Young had discovered the wave nature of light.  Doppler realized that as waves passed by an observer, motion towards the source would make the wave appear to have a higher frequency (shorter wavelength) and shift the color of the light towards the blue end of the spectrum.  By similar reasoning, an observer moving away from the source would appear to have a lower frequency, a longer wavelength, and therefore shift the color to the red end of the spectrum.  This seemed the perfect natural explanation for the color pairings of the binary stars.  The blue star was moving towards the Earth as it moved in its orbit and the red star was moving away from the Earth (see Figure~\ref{fig:doppler}).  It was a perfectly logical explanation that was completely wrong.

Measured values for the speed of light were already available from the work of Roemer and others, yielding values on the order of 300,000 kilometers per second (186,000 miles per second).   It was quickly realized that such a large color shift in either direction, on the order of a 20-50\% change in wavelength, would imply incredible velocities for the stars.  Application of Kepler's 3rd Law to these binary star systems, as well as angular measurements to determine the scales (the first successful stellar distance determinations had been made a few years earlier by Bessel in 1838)\footnote{Prior to this, stellar distances were often estimated photometrically, i.e. their distances were computing using the inverse-square law of light and assuming the stars were the same luminosity as the Sun\citep[pp. 7-8]{2006ccha.book.....L}.  This biased the distance estimates to lower than the actual value, since many of the bright stars we see at night we now know are intrinsically brighter than the Sun.} indicated such systems could not be gravitationally bound if the velocities were as high as Doppler suggested.  Many binary stars had already had their orbits mapped, and the results did not support Doppler's theory.

\subsection{Wrong, But Useful}
But all was not lost, for while it was quickly recognized that Doppler's theory would not explain the colors of double stars, others recognized that sound, which also propagated as waves, should have a similar property.   In 1845, Christoph Buijs Ballot successfully conducted the experiment with a group of trumpeters playing a single tone on a train traveling at the incredible (for its day) speed of 40 miles per hour.  This was a fairly leading-edge experiment which can today be conducted by anyone trying to cross a busy highway.  The experiment was possible because the speed of sound in air, about 760 miles per hour, was less than twenty times larger than the fastest speeds available, in this case, 40 miles per hour. The resulting change in pitch, about five percent, was detectable with the techniques of the day.  However, the speed of light is \textit{much} higher, so the change in frequency was still far too small to detect in the laboratory spectroscopes of the day.

A few years later, in 1848, Hippolyte Fizeau would independently propose the same mechanism for light, specifically suggesting measuring the displacement of spectral lines.  Fizeau recognized the importance of the spectral lines over a decade before the pioneering work of Bunsen and Kirchhoff, which officially launched spectroscopy as a science, though the idea that spectral lines revealed intrinsic properties of the stars dates back at least to William Herschel (see Section~\ref{sec:spectra}).

The first attempts to measure the Doppler effect in stellar spectra proved difficult and many early claims were questionable.  For stellar radial velocities, H.C. Vogel developed techniques in the late 1880s and into the 1890s which obtained the most robust measurements of many bright stars visible from his Potsdam observatory\citep{1889AN....121..241V, 1890MNRAS..50..239V}, including measurements which supported the eclipsing nature of Algol ($\beta$ Persei)\citep{1890PASP....2...27V}.  Around this same time, the rotation of the Sun would be measured by Dun\'{e}r spectroscopically\citep{1890AN....124..267D,1890PASP....2..192H}.  

In a summary of the progress in radial velocity determination as of 1900, Vogel would comment on the still open question in the physics community as to whether the Doppler principle actually applied to light.  He pointed to the agreement of Doppler measurements by \citet{1890AN....124..267D} with solar rotation measured by sunspot motion as a verification of the concept\citep{1900ApJ....11..373V}.  The accuracy of spectroscopes were still not yet up to the task of measuring the Doppler effect for light in an Earth laboratory!  

However, the following year, \citet{1901ApJ....13...15B}, who was also conducting radial velocity studies at the Pulkova observatory, would demonstrate a laboratory instrument which generated sufficiently high velocities to produce a detectable Doppler shift for a laboratory spectroscope.  It achieved high velocities through multiple folding of the light path between two moving mirrors.  This finally provided a laboratory verification of the Doppler effect for light.

\subsection{Doppler-Based Technologies}
For the past decade, the SOHO satellite has flown the Michelson Doppler Interferometer (MDI) that takes high-resolution dopplergrams of the solar disk on timescales as short as minutes\citep{1995SoPh..162..129S}.  In these images, each pixel value represents a radial velocity measurement.  These images are used to measure the vibrational modes of the Sun, useful for determining not only its internal structure, but enabling scientists to observe the formation of sunspots on the side of the sun \textit{not} facing the Earth\citep{2003soho...12...15B}.  This capability is important for space weather forecasting to detect sunspots and active regions before they come around the solar limb.  Such forecasting is critical for protecting astronauts in space as well as large-scale electrical grids on Earth and sensitive electronics on satellites.

The Doppler effect must be considered in almost any high-frequency application involving signal transmission between distant objects in relative motion, including satellites and interplanetary spacecraft.   More down-to-Earth applications include LIDAR (LIght Distance And Ranging), the equivalent of using lasers for distance measurement; the Global Positioning System (GPS), Doppler radar (which measures distance as well as velocity) used in tracking weather; and sonic medical imaging using the Doppler effect with sound.  Like gravity, the first insights for these technologies were developed as part of a problem in astronomy.

%% file: atoms2stars.tex
\section{From the Center of the Atom to the Center of the Stars}
\subsection{What Powers the Stars?}
One of the big questions in astronomy in the late 1800s had become the question of age of the Sun.   Intimately related to the question of the Sun's age was the question of its energy source.  Mathematical techniques and observational data, as well as physical understanding, had reached a level where astronomers and physicists were beginning to address these questions in detail.

Early calculations of the Sun's age were based on theories that it was powered by chemical energy, such as provided by coal burning in oxygen.  These calculations yielded life expectancies on the order of 6,000-10,000 years\citep{1854Thomson:fk}.  This was excellent news for those who believed the Biblical time scales, but inconsistent with data accumulating from other observations in geology and biology.  The other problem with the chemical fuel scenario was familiar to anyone who has tried to keep a fireplace burning -- how do you keep the accumulating ash from suffocating your fire?

In the latter part of the 1800s, work by William Thomson (Lord Kelvin) and Hermann von Helmholtz estimated the timescales for powering the Sun by gravitational collapse.  Gravitational potential energy would be converted into the thermal energy of the gas, which would radiate the energy and cool the gas.  This generated a longer age estimate, on the order of 20 million years, but was still inconsistent with evidence from geology and biology.  The meteoritic scenario proposed by Lockyer, where the Sun was powered by a continuous infall of meteoritic material, had similar issues.  Clearly there was a problem of physical consistency in the sciences that required a solution.  It would be the first hint of an undiscovered source of energy and the solution would lie in as yet undiscovered properties of the atom, which had been viewed as an indivisible particle since the time of the Greeks.

\subsection{The Atom becomes `Divisible'}
In 1897, J.J. Thomson proposed that the particle known today as the electron was the underlying cause of mysterious ``cathode rays'' which carried electrical currents in evacuated glass tubes.  This broke with the Greek notion that atoms were indivisible structures and initiated a new wave of experiments to probe that structure. 

The discovery of radioactive decay in 1901 by Ernest Rutherford, as well as the first experiments attempting to discern the structure of the atom, hinted at solutions to this dilemma.  But even these discoveries seemed to create more questions than answers, as experiments indicated the atomic nucleus was much smaller, yet much more massive than the surrounding electrons.  This created problems for Newtonian mechanics and Maxwell's electromagnetic theory, as attempts to model this structure mathematically suggested the electrons would radiate away their energy, causing atoms to collapse in microseconds.  

Niels Bohr's orbital model of the atom began to lead the way out of this quagmire\citep{1913Bohr:uq}, with its hints that the subatomic world could be very different from that of normal human experience.  The notion that the atom had an internal structure consisting of a massive positively charged nucleus surrounded by light, negatively charged electrons, entered the consciousness of the physics community.

While trying to reconcile the coordinate transformation properties of Newton's mechanics and Maxwell's electrodynamics, Albert Einstein would reformulate mechanics\citep{1905Einstein:xz}.  A by-product of this reformulation would be the famous mass-energy equivalence\citep{1905Einstein:jb} or
\begin{equation}
E = m c^2
\end{equation}
Today, this equation, and its association with Einstein, may be the world's most recognized equation from physics\citep{2001Bodanis}.

Between 1924 and 1926, Werner Heisenberg published his matrix formulation of quantum mechanics\citep{1925ZPhy...33..879H} and Erwin Schrodinger published his wave equation\citep{1926Schrodinger:kx}.  With these new tools, others quickly discovered that it explained the success of the Bohr model in describing the hydrogen atom\citep{1926PhRv...28..927E} while resolving many of its failures.  Shortly thereafter, Paul Dirac would successfully integrate special relativity with wave mechanics and the result suggested the existence of a world of anti-particles\citep{1928RSPSA.117..610D,1928RSPSA.118..351D}.  Dirac would initially propose that the positive proton was the antiparticle of the electron, in spite of their radically different masses\citep{1930RSPSA.126..360D}.

\subsection{The Convergence of Stellar Structure and Physics}
While some still held to the notion of stars powered by gravitational contraction, the theoretical work by Arthur Eddington, combined with the growing base of astrophysical data, would demonstrate that such a mechanism could be excluded due to the observed period stability of pulsating stars.  Theoretical models demonstrated that a pulsating star powered by gravitational collapse would exhibit a change in its pulsation period far larger than observed\citep{1918Obs....41..379E,1919MNRAS..79R.177E}.

Even without knowing details of the energy generation mechanism, many general features of stellar structure could be understood using the physics of the day: the gas laws, gravitation, and heat transfer.  These simplified models, called polytropes, could explain many basic stellar characteristics, such as central temperatures and pressures.  These parameters enabled physicists to integrate the knowledge of the stellar interior with their growing understanding of atomic structure and they then began proposing testable theories about the power source of the stars.  Two ideas became prominent in the 1920s and these were examined using the new theory of quantum mechanics by \citet{1931MNRAS..91..283W}\footnote{As a student of Ralph Fowler, this would be Alan H. Wilson's only paper on astrophysics.  He would gain wider recognition when he applied the new quantum theory to semiconductors and wrote the two papers which laid the foundation of modern semiconductor theory\citep{1931RSPSA.133..458W, 1931RSPSA.134..277W}.}:
\begin{itemize}
\item Direct mass-energy conversion  
\item Nuclear fusion, or building up of heavy nuclei from lighter ones.  
\end{itemize}
The arguments over which process was occurring were fought out in the scientific literature and both positions had its defenders.  

Sir James Jeans supported the mass-energy conversion process, or matter annihilation as it was sometimes called in the literature.  With Dirac's proposal of the existence of antimatter, and early speculations that the proton was the antiparticle of the electron, these concepts became integrated into Jeans' theory.

Arthur Eddington was one of the original proposers of the nuclear fusion process, specifically that four hydrogen nuclei could fuse to form one helium nucleus.  He would become one of the major advocates of this mechanism\citep{1920Obs....43..341E}\footnote{I've found a number of references attributing an early proposal of helium formation from fusion of hydrogen to William Harkins.  However, examination of some papers by Harkins between 1916-1920 reveal that while he did propose an atomic model where helium was constructed from hydrogen, I did not find a clear statement proposing stars as a location for this process\citep{1916PNAS....2..216H,1917Harkins:fk,1920PhRv...15...73H}.}.

But initial calculations were not encouraging.  Simple polytrope stellar models, incorporating gravity and the gas laws, enabled astronomers to estimate the temperature and pressure in the centers of stars.  It turned out that the temperature estimates for stellar cores, about 15 million K for a star the mass of the Sun, corresponded to thermal energies of hydrogen nuclei insufficient to overcome the coulombic repulsion of the positively-charged hydrogen nuclei.  In spite of this, Eddington continued to defend the idea.  Probably his most famous retort to critics can be found in his work \textit{``The Internal Constitution of the Stars''}:

\begin{quote}``For example, it is held that the formation of helium from hydrogen would not be appreciably accelerated at stellar temperatures, and must therefore be ruled out as a source of stellar energy. But the helium which we handle must have been put together at some time and some place.  We do not argue with the critic who urges that the stars are not hot enough for this process; we tell him to go and find a hotter place.''\citep[pg 301]{1926QB801.E21}\end{quote}

\subsection{Tunneling Deep into the Structure of the Atom}
Progress was also being made on other fronts, particularly in understanding the structure and interactions of atoms, that would impact the question of the stellar energy source.  Shortly after publication of the Schrodinger equation, several researchers realized that the fact it was a wave equation could give material particles other interesting wave-type properties.   One of these properties was the ability to not only be reflected but also \textit{transmitted} at an interface, or in the case of sub-atomic particles, at an energy barrier.  Due to the probabilistic nature of quantum mechanics, this would actually be a \textit{probabilistic} process which could be seen in the laboratory only with large numbers of particles or repetitive attempts.

In 1928, Fowler and Nordheim discovered that this transmission probability could explain the unusual process of electron emission from cold metals in high external electric fields, a process known as field-effect or cold-cathode emission\citep{1928Fowler:fk}\footnote{Cold-cathode emission was discovered in 1922 and had defied conventional explanations.  Unlike hot-cathode emitters, cold-cathodes did not require a heating filament to liberate electrons from the metal surface.}.  The result of their derivation of current/voltage characteristics for this process became known as the Fowler-Nordheim equation.  On other fronts, George Gamow\citep{1928ZPhy...51..204}, as well as R.~W. Gurney and E.~U. Condon\citep{1929PhRv...33..127G} would discover that this barrier penetration process could explain some features of alpha-decay, specifically the relationship between the half-life of the radioactive nucleus and the energy of the emitted alpha-particle.  \citet{1929ZPhy...54..656A} would propose this barrier penetration process, or ``tunneling'', could overcome the coulombic barrier penetration problem that hindered the proton-to-helium process.  Unfortunately, there were still a few mysteries in the atomic nucleus that hindered computing an actual solution to the problem.

In 1931, a positively-charged particle with the mass of the electron was discovered in showers of cosmic rays\footnote{There is a report that the positron was seen by other means as early as 1930, but the researcher did not recognize the significance.}.  Astrophysics had again provided a laboratory not yet made available by technology.  This particle was identified as the anti-electron of the Dirac theory and proved to be fatal to Jeans' idea that stars were powered by electron-proton annihilation.  The positron would later be discovered to be a nuclear decay channel.

The following year, James Chadwick discovered the nucleus also contained a heavy neutral particle, with a mass near that of the proton.  With this discovery, the mystery of atomic isotopes, atoms with the same chemical properties but different masses, was solved.  All the constituents of normal atoms were now known.  The next step was now filling in the details of how the atomic constituents interacted.

Fusion of light nuclei into heavier nuclei was demonstrated in 1934\citep{1934Oliphant:ly}.  Some confusion  surrounded the discovery of nuclear fission of the atomic nucleus.  It was believed Enrico Fermi may have achieved it as early as 1934\citep{1934Fermi:kx} but it was not recognized as a fission reaction until 1939\citep{1939Meitner:uq}, after the work of Otto Hahn and Fritz Strassmann made a conclusive experiment\citep{1939Hahn:fk}.  While these discoveries hinted at the amount of energy locked up in the mass of the atomic nucleus, they were still confined to table-top, or perhaps room-sized, experiments.

\subsection{Into the Core of the Stars}
Also in the 1930s, cyclotron particle accelerators were beginning to operate at energies equivalent to those estimated in the cores of stars.   Soon, the idea of barrier penetration were being tested for the nucleus and confirming the earlier theoretical work.   Gamow and Edward Teller would soon examine the physics of these reactions taking place under high temperatures like those in the center of stars\citep{1938PhRv...53..595G,1938PhRv...53..608G}.  

When Gamow organized a convention of physicists and astrophysicists in 1938, the two major components of the problem: the understanding of stellar structure, and the understanding of atomic structure, were finally in place.  The physicists had the results of their theories of the atomic nucleus and small-scale experiments.  The astrophysicists had very good ideas about the composition and structure of the stars, much of it derived without knowing the specific energy source, beyond it being located near the center of the stars\citep{1967Bethe:sh,1968Bethe:fh}.  Later that year, Hans Bethe would solve the problem which had eluded Eddington and his critics.

Bethe computed the theoretical reaction rate for two protons to fuse into a deuteron\footnote{The deuteron, also designated $^2H$ is an isotope of hydrogen consisting of one proton and one neutron.} in the core of the Sun.  Instead of considering only the coulombic repulsion in the reaction, he would incorporate the barrier penetration process from quantum mechanics\footnote{Bethe did not explicitly include the neutrino in his original analysis, though he did use the Fermi\citep{1934ZPhy...88..161F} and Gamow-Teller\citep{1936PhRv...49..895G, 1937PhRv...51..289G} theories for computing the $\beta$-decay probability of positron emission which implicitly included it.}:
\begin{equation}
\mathrm{{\rm ^1H} + {\rm ^1H} \rightarrow {\rm ^2H} + e^+ + \nu_e}
\end{equation}
The computed barrier penetration probability was extremely low, at the solar center temperature and density, the mean lifetime of a proton was computed to be about 10 billion years\citep[p. 369]{1983psen.book.....C}.  In Bethe's time, the estimates of temperature, composition, and density in the solar core were approximate, but still surprisingly close to values determined by more refined modern techniques.  Using those numbers, he obtained a value for the total energy production of the Sun, 2.2 ergs/gm/sec, the same order of magnitude of that observed from the Sun of 2.0 erg/gm/sec\footnote{At the time, many astronomers believed the Sun was composed largely of iron, based on the work of \citet{1914Russell:fk}.  Work by \citet{1925PNAS...11..192P}, \citet{1932MNRAS..92..471E}, and \citet{1929ApJ....70...11R} already indicated the stellar atmosphere was mostly hydrogen, but it wasn't until the post-WWII years that Hoyle conclusively demonstrated that the bulk composition of stars had to be hydrogen due to opacity constraints\citep{1946MNRAS.106..255H}.}.  For comparison, the human body generates about 150 watts so for a person weighing 150 kilograms, this energy generation is roughly 1 joule/kg/sec = 10,000 erg/gm/sec.  The energy density of the human body is far higher than the energy density of matter in the Sun, but bear in mind the Sun has \textit{much} more matter\footnote{Thanks to a Balticon (\url{http:\\www.balticon.org}) attendee for pointing out this interesting datum to me.}.   

With followup work, parameters in Bethe's calculation would be improved and Bethe himself would explore even more options for thermonuclear reactions possible in the stellar interior\citep{1939PhRv...55..434B}.  That same year, Carl von Weis{\"a}cker in Germany would reach similar conclusions about stellar energy sources\citep{1937Weizsacker:kx}.

\subsection{Igniting Stellar Energy on the Earth}
In August 1939, the first of the ``Einstein Letters'' (actually penned by Leo Szilard) was sent to Franklin Roosevelt, advocating the development of the atomic bomb.  At this time, there were no \textit{laboratory} experiments indicating nuclear reactions, such as the fission of uranium, could proceed at the temperatures and densities needed to produce an explosion.  The closest thing to a `data point' indicating that the laboratory understanding of the atomic nucleus could be extrapolated to the necessary temperatures and densities was the agreement with stellar energy sources obtained using the exact same physics. 

It is interesting to note that Carl von Weis{\"a}cker, the nuclear astrophysicist mentioned above who had paralleled some of the work by Bethe, is explicitly mentioned in some of the ``Einstein Letters''.  Von Weis{\"a}cker's close association with the German government was emphasized, as he was someone who would know that such a weapon was possible, raising concerns that the Germans might already be working on a nuclear weapon.

The atomic bomb, first detonated on July 16, 1945, would become the first demonstration of energy release by nuclear reactions under conditions similar to those in stars.  After WWII, Bethe and Teller, both doing nuclear astrophysics before the war, would become major players in the American nuclear weapons program, as would many other nuclear astrophysicists\citep{1972cht..conf..169K}.  Seven years later, November 1, 1952, the first hydrogen bomb detonation would use nuclear reactions first explored to explain the energy source of the stars\citep{1995Rhodes:uq}[pg 416-419].

\subsection{More Tunneling Applications}
Who could imagine that the quantum tunneling process, which was important for understanding the first stage in the energy generation process of the stars would find its way into technologies that we use every day?  Experiments with cold cathodes would continue throughout the 1920s and 1930s, with Philo T. Farnsworth submitting a patent for a cold-cathode electron discharge tube in 1936, which was granted in 1939, patent number 2,184,910\footnote{Farnsworth would be granted a second cold-cathode device patent in 1941.}$^,$\footnote{This name might seem familiar as Farnsworth invented many devices in the era of the electron tube.  He is also regarded as the inventor of television.}.  The old technology of radio tubes is an example of a hot-cathode technology and they are still used in some high-power applications.  Tubes based on the cold-cathode concept are all around us, in the form of fluorescent light blubs.   

In the late 1950s, Leo Esaki would successfully demonstrate quantum tunneling in solids, specifically semiconductors, with the invention of the tunnel diode\citep{1974Esaki:fk,1976Esaki:fk}.  Tunnel diodes are major components of semiconductor electronic devices.

Quantum tunneling has proven to be a double-edged sword in the field of microelectronics.  While it makes some new devices possible, it also created some problems.  The electron paths in modern VLSI (Very Large Scale Integration) circuitry are etched so close together, that the effects of quantum tunneling must be included in their design.  In some cases, the tunneling is part of the desired behavior for the circuit, but tunneling also creates leaks in the current flow that are a major source of heating in these devices.  Quantum tunneling using the Fowler-Nordheim equation, from the original work in cold-cathode emission, is also important in the operation of the flat-panel displays used in modern computers and high-definition television.


%% file: carbon12.tex
\section{What the Universe Taught Us About $^{12}C$}

\subsection{The Golden Age of Nuclear Astrophysics}
\begin{figure}[htbp]
\includegraphics[scale=0.8]{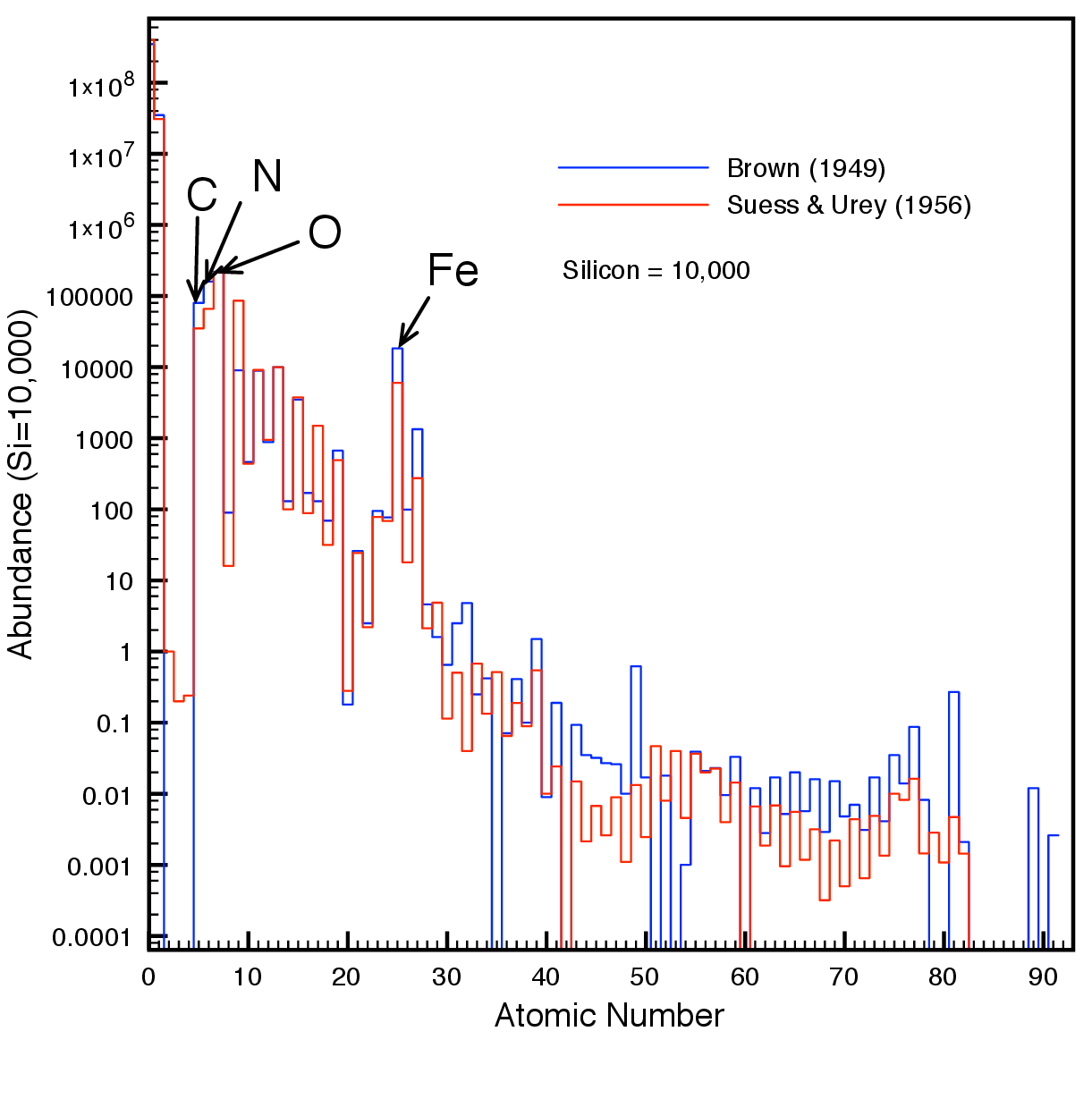}
\caption{Cosmic abundances by atomic number based on the data presented in \citet{1949RvMP...21..625B} and \citet{1956RvMP...28...53S}.  Values are normalized based on silicon=10,000.}
\label{fig:abundance}
\end{figure}

Prior to 1950, due to the growth in understanding of the structure of the atom and its implications for spectroscopy, astronomers were finally obtaining data of sufficient quality to determine the chemical composition of the cosmos.  They did these analyses using combinations of data from the Earth, meteorites, and solar spectroscopy.  Reliable measurements of elemental abundances were becoming available\citep{1949RvMP...21..625B, 1956RvMP...28...53S}, samples of which are plotted in Figure~\ref{fig:abundance}.

On the theoretical front, many of the light element nuclear reactions for stellar interiors had been explored by Bethe prior to 1939\citep{1938PhRv...54..248B,1939PhRv...55..103B}.   From 1940 to 1945, published research in stellar nuclear physics is almost non-existent as nuclear astrophysicists applied their knowledge towards the pursuit of national goals.  But after the close of the World War II, that would begin to change\citep{1972cht..conf..169K}.

The 1950s might be considered something of a ``Golden Age'' for stellar nuclear astrophysics.  The success of the Manhattan Project, and the growing state of international rivalry which would become the Cold War, kept research on the properties of the atom and the atomic nucleus well-funded.  Astrophysicists made extensive use of this data and began to explore a broader range of nuclear reactions that could take place at the high temperatures and densities of stellar interiors.  Much of this research would provide additional feedback and guidance to the nuclear laboratories.

This time period would also see a convergence of two lines of exploration, connecting questions of the age of the cosmos with the nucleus of the atom.   It would also yield an incredible insight on the origin of life in the Universe.

General Relativity had provided the first mathematical structure whereby cosmology could be treated as a real science.  Two major cosmological models emerged from this: expanding universe cosmology\citep{1931MNRAS..91..483L, 1931MNRAS..91..490L}, known today as the `Big Bang', and Steady-State cosmology\citep{1948MNRAS.108..372H}.  Both cosmologies started with a universe of predominantly hydrogen but were clearly unable to explain the production of the heavier elements\citep{1942ApJ....95..288C}, a shortcoming recognized even by their proponents\citep{1946PhRv...70..572G,1947PhRv...71..273G}.  But where else in the Universe would temperatures be sufficiently high for nuclear reactions to build the heavier elements?  The cores of stars was the only other location which came close to meeting the temperature and density requirements.

But there was a problem, first recognized by Hans Bethe back in his 1939 paper\citep{1939PhRv...55..434B}.  If one tried to build elements heavier than helium by capture of protons or helium nuclei ($\alpha$-particles), the lack of a stable nucleus with eight nucleons created a bottleneck.  You could create reactions to build heavier nuclei:
\begin{eqnarray}
{\rm ^{4}He} + 2 {\rm ^1H} & \rightarrow & {\rm ^6Be} \\
2  {\rm ^{4}He} + {\rm ^1H} & \rightarrow & {\rm ^9B} \\
2  {\rm ^{4}He} & \rightarrow & {\rm ^{8}Be} \\
{\rm ^{8}Be} + {\rm ^1H} & \rightarrow  & {\rm ^9B} 
\end{eqnarray}
but once created, they would quickly disintegrate either spontaneously or with the very next reaction:
\begin{eqnarray}
{\rm ^9B} + {\rm ^1H} & \rightarrow  & {\rm ^{8}Be} + {\rm ^2H}  \\
{\rm ^{11}B} + {\rm ^1H} & \rightarrow & 3 {\rm ^{4}He} \\
{\rm ^{8}Be}  & \rightarrow & 2  {\rm ^{4}He}
\end{eqnarray}
All combinations of two-nuclei reactions invariably produced ${\rm ^{8}Be}$, and occasionally a lighter nucleus.  The ${\rm ^{8}Be}$ would then quickly decay into two $\alpha$-particles in $\sim {10}^{-\mit  16}$ seconds.

The only way out of this dilemma was to step up from two-nuclei reactions to three-nuclei reactions.  Bethe proposed that three helium nuclei could fuse to form ${\rm^{12}C}$, which would bridge the instability gap, providing a stable nucleus from which heavier elements could be built by captures of hydrogen and helium nuclei.  The reaction is often referred to as the triple-alpha reaction.
\begin{equation}
{\rm ^{4}He} + {\rm ^{4}He} + {\rm ^{4}He} \rightarrow {\rm ^{12}C}
\end{equation}

Initially, this seemed to solve the problem of building the elements heavier than helium, but a new problem would quickly emerge as astrophysicists began to compare the amounts of carbon and heavier elements produced by these calculations with the observationally determined abundances of these heavier elements.  Not enough carbon was being produced, and this affected abundances of the heavier elements as well\citep{1950RvMP...22..153A}.

\subsection{Cooking Helium}
Fred Hoyle, an advocate of Steady-State cosmology, realized that the carbon abundances had to be produced in the stars, lest there be no astrophysicists around in the future to even ponder the question.

Hoyle decided to re-examine the carbon formation problem.  As noted by Bethe, two ${\rm ^{4}He}$ nuclei could fuse to form a ${\rm ^8Be}$ nucleus, but this would decay back to two ${\rm ^{4}He}$ nuclei in about $10^{-16}$ seconds.  Most astrophysicists regarded this as a problem in the realm of temperatures (about 20 million K) they originally examined, because there was not enough time for the third ${\rm ^{4}He}$ nucleus to fuse.  They could generate the appropriate amount of carbon at much higher temperatures (about a billion K) but the physics just didn't support the possibility of stellar cores reaching that temperature at this stage of their evolution.  Hoyle re-examined the reaction at higher densities and found that the reaction sequence
   \begin{eqnarray} 
         2  {\rm ^{4}He} & \rightarrow & {\rm ^{8}Be} \\
      {\rm ^{8}Be} + {\rm ^4He} & \rightarrow & {\rm ^{12}C} 
   \end{eqnarray}
could proceed at even higher densities and temperatures, but still at a very low rate.  But this reaction still did not produce enough carbon.  

Other researchers had noted that the energy level structure of carbon was poorly known, an issue that could dramatically affect the reaction rate calculations\citep{1953ARNPS...2...41A}.  Hoyle made a bold proposal, that there was indeed a resonance in the energy-levels of the carbon nucleus which could bring the reaction rate up to a level needed to produce the necessary relative amounts of carbon (and oxygen) in stellar interiors.  Working backwards, he reformulated the reaction equations incorporating the unknown level and proceeded to compute where the level needed to be to produce the observed ${\rm ^{12}C/^{16}O}$ abundances.  He analysis predicted an energy level at 7.7MeV\footnote{MeV = million electron volts.  A unit of energy commonly used in particle physics.} above the ground-level energy state of the carbon nucleus\citep{1954ApJS....1..121H}.  

Hoyle discussed the carbon formation problem with William Fowler at CalTech.  They visited the nearby Kellogg laboratory and asked the researchers whether this energy level had been observed.  The researchers noted there had been some unconfirmed reports of a level near that energy, so they decided to set up an experiment to test it further.  They found the energy level at 7.68 MeV, precisely where Hoyle had predicted\citep{1953PhRv...92..649D}\footnote{Note that while the discovery paper predates the prediction paper, this is really an artifact of the research time and publication timelines.  Hoyle developed the analysis and recognized the problem, solved it, and then completed the paper.  The ${\rm^{12}C}$ resonance discovery paper\citep{1953PhRv...92..649D} credits Hoyle for pointing out the astrophysical significance of the level.}.  An excellent popular-level description of this discovery is available in Hoyle's autobiography\citep[chapter 16]{1994hiww.book.....H}.

This was probably the first (and only?) actual discovery based on what is today known as the ``Weak Anthropic Principle''.  The Weak Anthropic Principle might best be stated as the history we discover about the Universe will be consistent with the formation of carbon-based life today.  If one advocates a supernatural or non-naturalistic process for the evolution of the Universe, this energy level does not need to exist.  Nonetheless, it provided a compelling link between the human species and the Cosmos that Carl Sagan would express it in the statement ``We are made of starstuff''\citep[pg 233]{1980Sagan}.  The idea would even find its way into popular music:
\begin{verse}
\textit{We are stardust, we are golden, \\
We are billion year old carbon, \\
And we got to get ourselves back to the garden.\\}
\indent --- ``Woodstock''.  Written by Joni Mitchell.  Performed by Crosby, Stills, Nash and Young\footnote{Thanks to Stan Woosley for pointing this out at ``Astronomy with Radioactivities V'', Clemson University.  September 2005.}.
\end{verse}

%% file: implications.tex
\section{Implications and Consequences}

When it comes to discoveries in fundamental science, few of the discoverers have any inkling of the eventual consequences of their discoveries.  This is especially true in understanding the connections between science and technology.  None of the physicists who worked to understand the structure of the atom realized the technology impact their work would have.

When Newton imagined firing a cannonball around the Earth, did he picture the practical benefits of the capability, Earth-orbiting satellites, as it is used today, nearly 300 years in the future?  Did he imagine the physics he founded would be used to navigate spacecraft to places which were only small disks in a telescope to him?  I suspect he did not.  The notion of travel to other worlds was still the realm of fiction.  Yet today we take for granted technologies available due to his insight.

This illustrates the power of deductive reasoning in science which can enable scientists to make giant leaps forward in understanding.  If Newton had reasoned empirically based on the experiments possible in his day, such as the work of Galileo, there would have been no reason to believe that gravity followed an inverse-square force law.  Even today, measuring an inverse-square law of gravitational forces at laboratory scales is a \textit{very} difficult experiment.  Yet, by deducing a property of gravity, developing the consequences, and then comparing those consequences to observations available in Nature, human understanding moved forward by leaps and bounds.

Some members of the school of empiricist thought like to argue ``what if you choose the wrong theory?''  We can answer this question with examples from the history of science.  In the case of Newtonian gravity, we have an example in the discovery of the anomalous perihelion shift of Mercury\footnote{This is the discrepancy discovered by LeVerrier described in Section~\ref{sec:gravity}.}.  It would take over fifty years to solve that mystery.  Did that mean Newton was wrong?  No.  Every scientific theory has its domain of applicability, every theory has realms where their approximations work, and realms where their approximations break down.  We don't use Newtonian gravity to build buildings on the Earth (unless the building is \textit{very} tall), we use Galileo's model of gravity.  We don't use Einstein's theory of gravity for navigating the space shuttle when Newton's theory works to the level of precision needed for the task.  The relevant question is ``Could we have learned the greater understanding revealed by Einstein without the two centuries of observations, analysis, and \textit{experience} developed under Newton's ideas?''  I think the answer is probably ``no''.

To be fair, some of these discoveries probably would have been made without the intervention of astrophysics.  Many were on the verge of being technically possible and the astrophysical observations provided an additional incentive to examine them more closely.  But it was the astrophysical problem of gravity and the structure of the solar system that opened the door to the exploration.  In these cases we have seen that not all science is ``extrapolated'' from the Earth into the distant cosmos, but in fact a significant amount is ``interpolated'' from cosmic observations into applications close to the Earth.

The science you know determines the technology you can achieve, and any modern technology often requires the integration or synthesis of multiple components of a science.  If any one of the scientific components is missing or wrong, the technology doesn't work. 

In the second paper of this series, I'll explore some other cosmic science that would take years to become testable in Earth laboratories, and visit some technologies that were almost unsuccessful, due a failure to understand some important cosmic science.